\documentclass[3p,times,procedia]{elsarticle}
\usepackage{nupha_ecrc}

\volume{00}

\firstpage{1}

\journalname{Nuclear Physics A}

\runauth{Ranbir Singh}

\jid{nupha}

\jnltitlelogo{Nuclear Physics A}

\usepackage{amssymb}

\usepackage{lineno}

\usepackage[figuresright]{rotating}

\newcommand{\snn}{$\sqrt{s_{\mathrm{NN}}}~$}
\newcommand{\s}{$\sqrt{s}~$}
\newcommand{\kst  }{$\mathrm{K^{*0}}~$}
\newcommand{\ks}{$\mathrm{K^{0}_{S}}~$}
\newcommand{\pt}{$p_{\mathrm{T}}~$}
\newcommand{\rh}{$\rho_{00}~$}

\begin{document}

\begin{frontmatter}



\dochead{XXVIIth International Conference on Ultrarelativistic Nucleus-Nucleus Collisions\\ (Quark Matter 2018)}

\title{Spin alignment measurements using vector mesons with ALICE detector at the LHC}


\author{Ranbir Singh (For the ALICE Collaboration)}

\address{School of Physical Sciences, National Institute of Science Education and Research, HBNI, Jatni-752050, India}

\begin{abstract}
We present new measurements related to spin alignment of \kst vector mesons 
at mid-rapidity for Pb--Pb collisions at \snn = 2.76 and 5.02 TeV.  
The spin alignment measurements are carried out with respect to
production plane and second order event plane. At low \pt the spin
density  matrix element \rh for \kst is found to have values
slightly below 1/3, while it is consistent with 1/3, i.e. no spin
alignment, at high \pt. Similar values of  \rh  are observed
w.r.t. both production plane and event plane. Within statistical and
systematic uncertainties, \rh  values are also found to be
independent of \snn. \rh also shows centrality dependence with maximum
deviation from 1/3 for mid-central collisions w.r.t. both the
kinematic planes. The measurements for \kst in pp collisions at \s = 13
TeV and for  \ks (a spin 0 hadron) in 20-40\% central Pb--Pb collisions at \snn =
2.76 TeV  are consistent with no spin alignment.

\end{abstract}


\end{frontmatter}


\section{Introduction}
\label{intro}
Relativistic heavy-ion collisions are expected to produce quark gluon
plasma with large
angular momentum~\cite{Becattini:2007sr} and intense magnetic
field~\cite{Kharzeev:2007jp}. One of the main goals of the ALICE physics
program in heavy-ion collisions
is to look for the signatures of these effects. The proposed
signature is the measurement of the spin alignment of vector mesons. The spin
alignment measurements can be performed by studying the angular
distributions of the vector meson decay
daughters~\cite{Abelev:2007zk,STAR:2017ckg,Abelev:2008ag}. 
The angular distribution of vector mesons~\cite{Schilling:1969um} is given by 
\begin{equation}
\frac{\mathrm{d}N}{\mathrm{d}\cos{\theta^{*}}} = N_{0} [ (1 - \rho_{00}) + \frac{1}{R}\cos^{2}{\theta^{*}}(3\rho_{00} - 1 ) ],
\label{eq:eq1}
\end{equation}
where $N_{0}$ is a normalization constant and $R$ is the event plane resolution. In
case of the production plane analysis $R=1$. \rh is the zeroth element
of the spin density matrix, which is a 3x3 hermitian matrix with unit
trace. Since \kst decays via the strong interaction, the diagonal
elements $\rho_{11}$ and $\rho_{-1-1}$ are degenerate and the only
independent observable is \rh. $\theta^{*}$ is the angle formed by one 
of the vector meson decay daughters in the rest frame of the vector meson with  the quantization 
axis or  polarization direction. The quantization axis can be 
normal to the production plane, that is determined by  the momentum of the vector meson and 
the  beam direction. It can also be normal to the reaction plane that is determined by the impact parameter 
and the beam direction.  The
polarization effects caused by either initial conditions or final state
effects would lead to non-uniform angular distributions of vector
mesons. This results in a deviation of the \rh from 1/3, which
indicates a net spin alignment whereas \rh = 1/3 signals no spin alignment. 

Here we present the new results related to the spin alignment of \kst
vector mesons in Pb--Pb collisions at \snn = 2.76 and 5.02 TeV measured
as a function of \pt and centrality. The spin
density matrix element \rh is measured w.r.t. the production and
event planes using the ALICE detector~\cite{Abelev:2014ffa}.

\section{Analysis details}
\label{sec:sp}
In this work we have measured the spin alignment of \kst at
midrapidity ($|y| < 0.5$) in Pb--Pb collisions at \snn = 2.76 and 5.02
TeV in various collision centrality classes.  The analysis is carried
out using 14 M minimum bias events in
Pb--Pb collisions at \snn = 2.76 TeV collected in 2010 (Run I) and 30 M
minimum bias events in Pb--Pb collisions at \snn = 5.02 TeV collected in 
2015 (Run II). The spin alignment results for \kst in pp collisions at \s = 13
TeV, as a baseline study, and of \ks in 20-40\% central
Pb--Pb collisions at \snn = 2.76 TeV, as null hypothesis, are reported
in~\cite{BM:sqm}. The spin alignment study of \kst is performed
w.r.t. the production and event planes, but the event plane analysis is
performed in Pb--Pb collisions at \snn = 2.76 TeV only.
 The detectors that are used for this analysis are Time Projection
Chamber (TPC) and Time-Of-Flight (TOF) detector for
particle identification at mid-rapidity and the forward detector V0
for triggering and centrality estimation. The V0 detector is also used
for the estimation of the $2^{nd}$ order event plane.
The \kst vector meson  is reconstructed through the invariant mass of its
decay daughters from its dominant hadronic decay channel (\kst$
\rightarrow \mathrm{K}^{\pm}\pi^{\mp}$). The major contribution to the
background in the invariant mass
distributions is combinatorial. It is estimated using
the mixed event technique, where opposite charged K and $\pi$ from
different events are mixed. These events are required to have similar
multiplicity and collision point position. Even after subtracting
combinatorial background still a certain amount of residual
background remains under the resonance peak that is described by
polynomial of $2^{nd}$ order in this analysis.
\begin{figure}
\begin{center}
\includegraphics[clip,scale=0.26]
{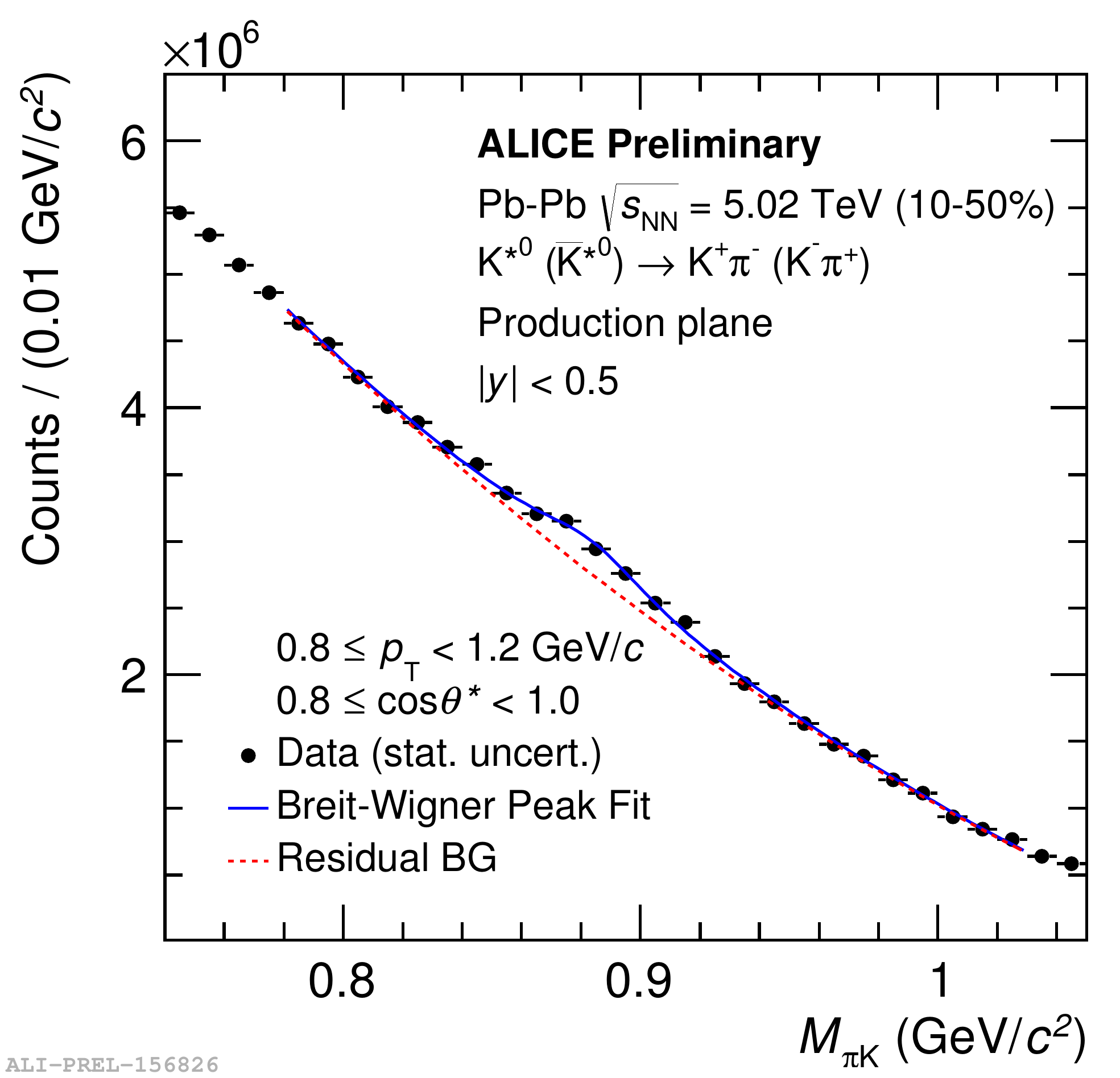}
\includegraphics[clip,scale=0.26]
{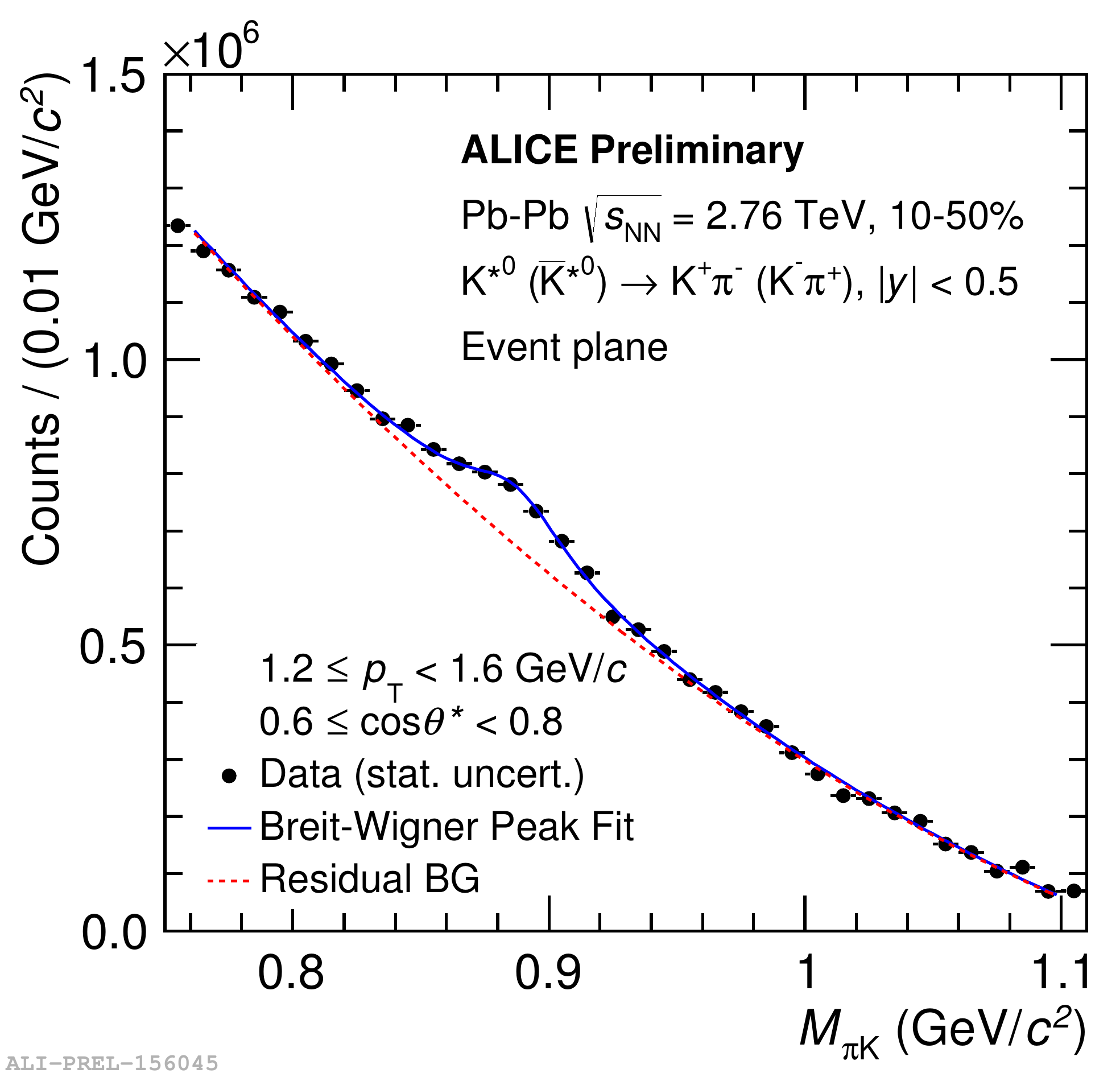}
\caption{(color online) Left panel: Invariant mass distribution
 of $\pi$K pairs at mid-rapidity after combinatorial background subtraction
 w.r.t. the production plane for 10-50\% central Pb--Pb collisions at \snn =
 5.02 TeV. Right panel:  Invariant mass distribution of $\pi$K pairs at mid-rapidity after combinatorial background subtraction
 w.r.t. the event plane in 10-50\% central Pb--Pb collisions at \snn =
 2.76 TeV. The invariant mass distributions are fitted with a
 Breit-Wigner function for the signal and a $2^{nd}$ order polynomial
 function in $M_{\mathrm{\pi K}}$ 
for the residual background. The error bars are statistical only.}
\label{fig:inv}
\end{center}
\end{figure}

\begin{figure}
\begin{center}
\includegraphics[clip,scale=0.26]
{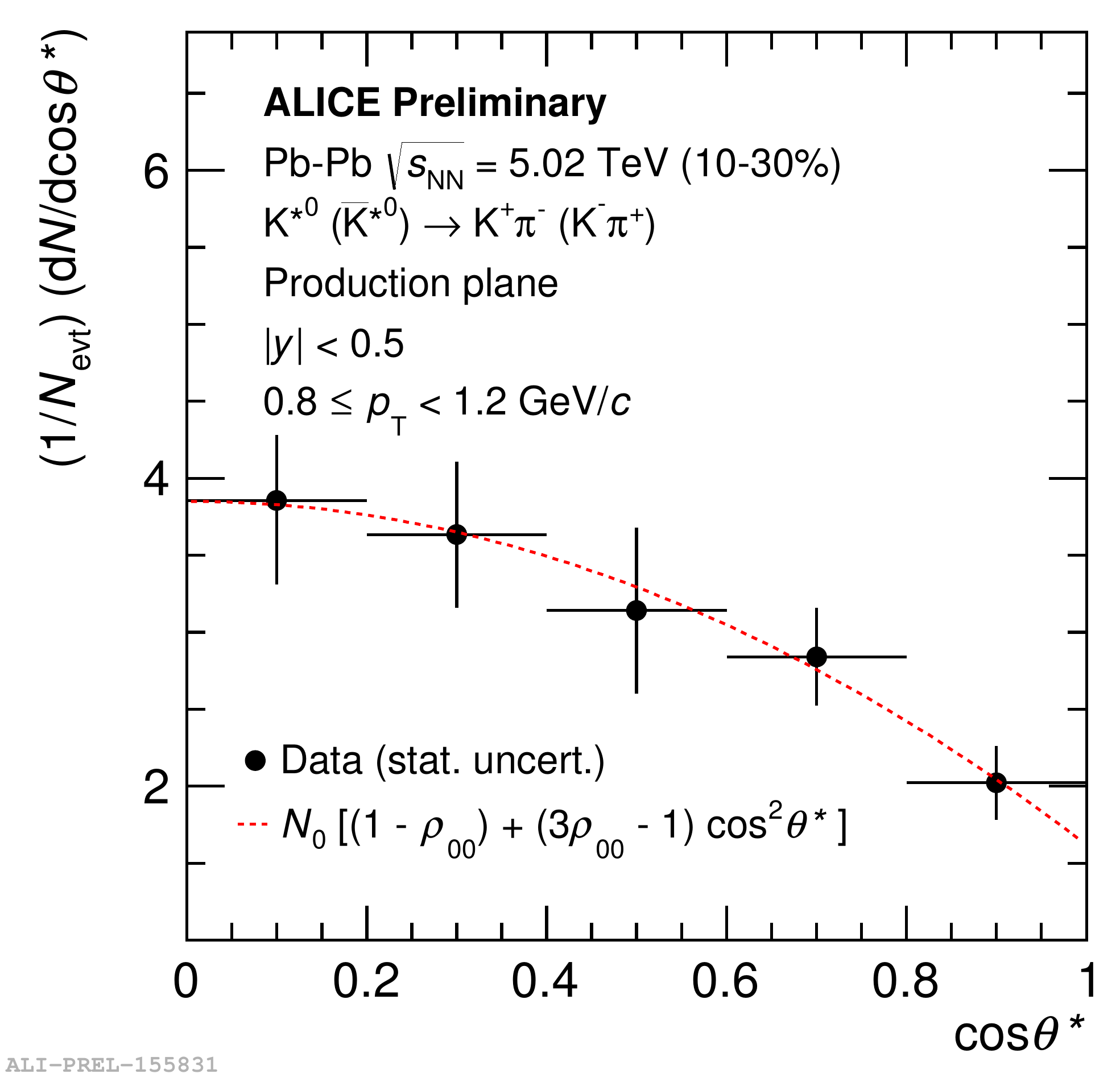}
\includegraphics[clip,scale=0.26]
{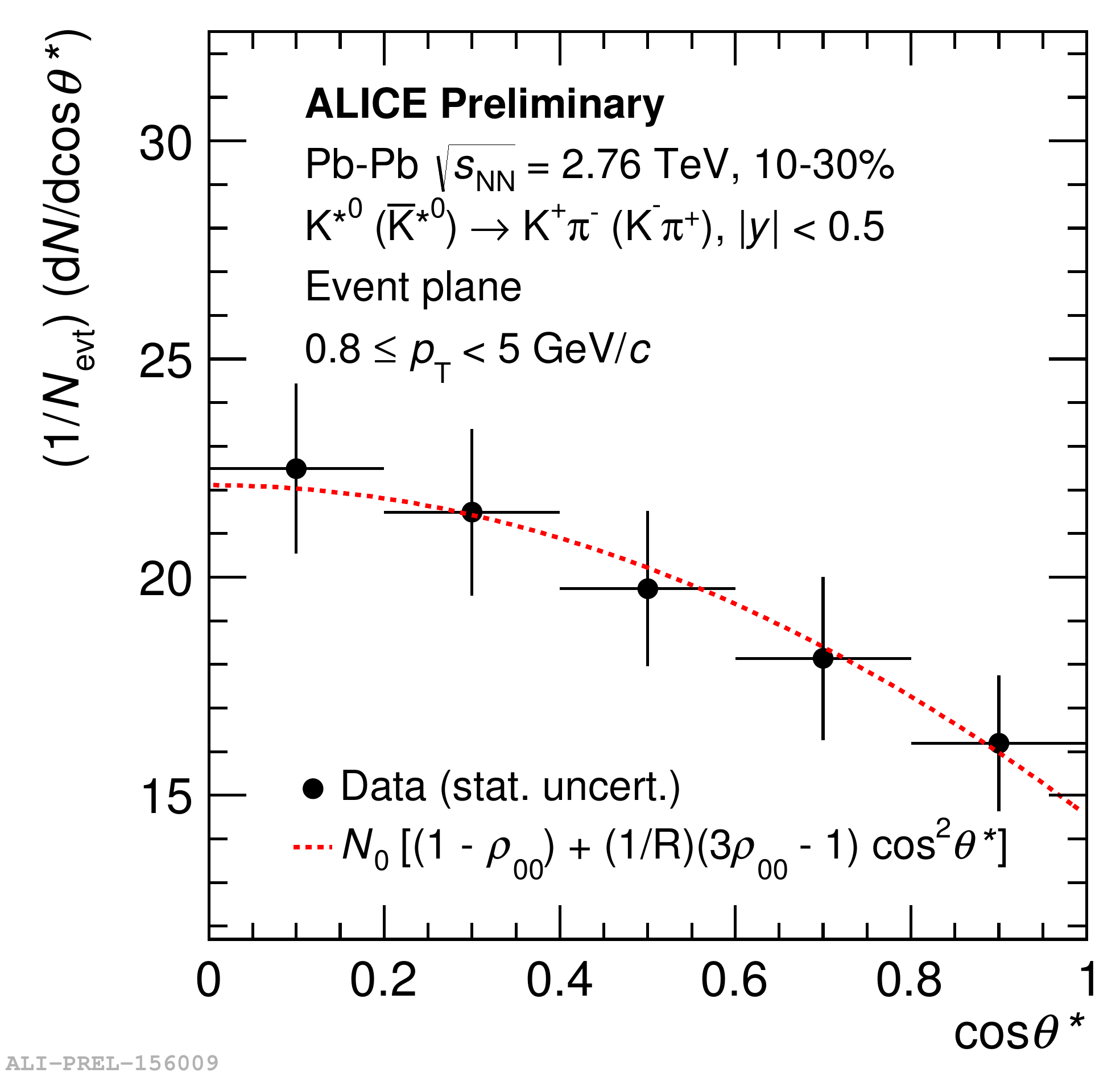}
\caption{(color online) Left panel:  $\mathrm{d}N/\mathrm{d}\cos{\theta^{*}}$ vs. $\cos{\theta^{*}}$ 
  at mid-rapidity w.r.t. the production plane for 10-30\% central Pb--Pb collisions
 at \snn =  5.02 TeV. Right panel:  $\mathrm{d}N/\mathrm{d}\cos{\theta^{*}}$ vs. $\cos{\theta^{*}}$ 
  at mid-rapidity w.r.t. the event plane
 for 10-30\% central Pb--Pb collisions at \snn = 2.76 TeV. The error bars are statistical only.}
\label{fig:costheta}
\end{center}
\end{figure}

The invariant mass distributions of $\pi$K pairs, after combinatorial
background  subtraction, are shown in Fig.~\ref{fig:inv} for Pb--Pb collisions.
These are the
representative plots shown for  a particular \pt and
$\cos{\theta^{*}}$ bin and the same procedure is followed for each
\pt~and $\cos{\theta^{*}}$ bin of the analysis. These distributions are
then fitted with a Breit-Wigner function for the signal and
a second-order polynomial for the residual background to extract the
\kst yield in each \pt and $\cos{\theta^{*}}$ bin
of a particular centrality class.
The \kst yields are then corrected for the corresponding  
reconstruction efficiency and acceptance evaluated using dedicated
Monte Carlo simulations~\cite{Adam:2017zbf,Abelev:2014uua}. 
The left panel of Fig.~\ref{fig:costheta} shows the
$\mathrm{d}N/\mathrm{d}\cos{\theta^{*}}$ distribution at
mid-rapidity, corrected for efficiency and acceptance, for 0.8 $\le$
\pt~$<$ 1.2 GeV/$c$   in 10-30\% central
Pb--Pb collisions at \snn = 5.02 TeV using the production plane. The
right panel shows
the same distribution for  0.8 $\le$ \pt~$<$ 5.0 GeV/$c$  in 10-30\% central
Pb--Pb collisions at \snn = 2.76 TeV using the event plane. The red dotted lines in the plots
are fits with the functional form reported in Eq.~\ref{eq:eq1}.
The \rh values for each \pt ~bin in various
centrality classes are extracted from the fits. 

\begin{figure}
\begin{center}
\includegraphics[clip,scale=0.27]
{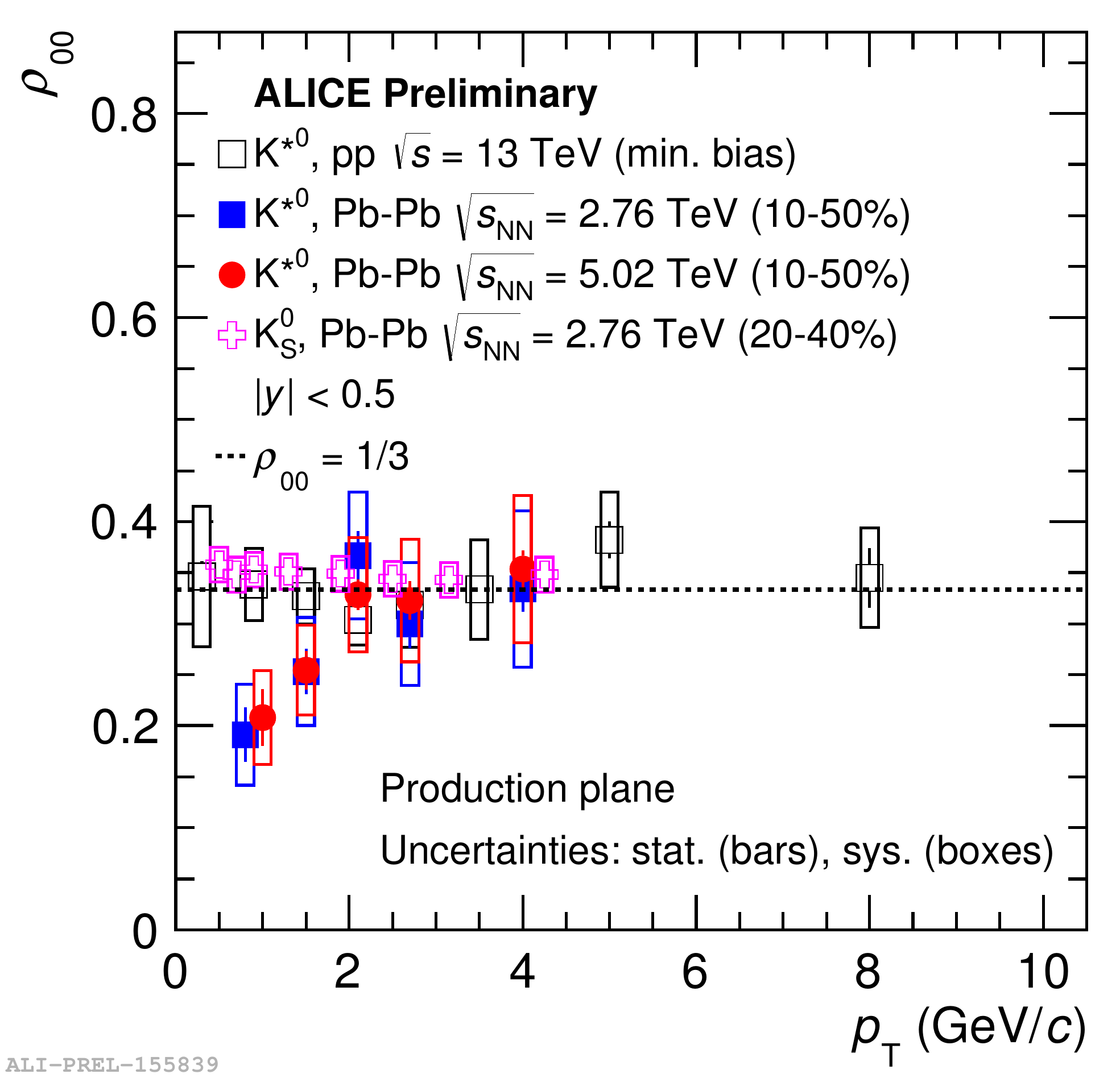}
\includegraphics[clip,scale=0.27]
{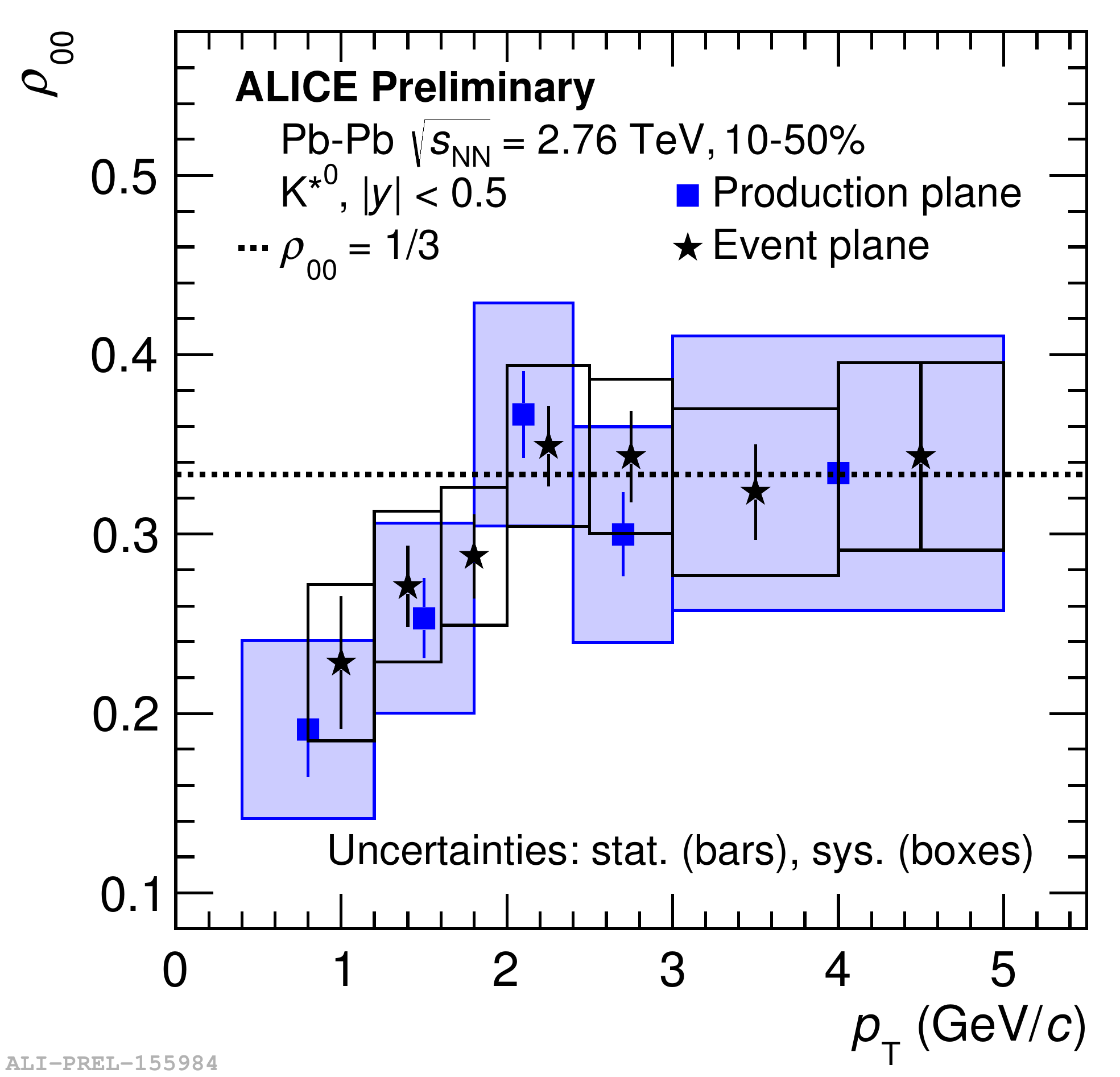}
\caption{(color online) Left panel: \rh vs. \pt of \kst w.r.t. the production
 plane in pp collisions at \s = 13 TeV and for 10-50\%  Pb--Pb collisions at
 \snn = 2.76 and 5.02 TeV. The corresponding results
 for \ks, a spin zero hadron, for 20-40\% Pb--Pb collisions at \snn = 2.76
 TeV are also shown.  Right panel: Comparison of
 \rh w.r.t. production and event planes in Pb--Pb
 collisions at \snn = 2.76 TeV. The statistical uncertainties
 are shown as bars and systematic uncertainties are shown as
 boxes. The dotted line at \rh = 1/3
shows the no spin alignment scenario.}
\label{fig:ptDep}
\end{center}
\end{figure}

\begin{figure}
\begin{center}
\includegraphics[clip,scale=0.27]
{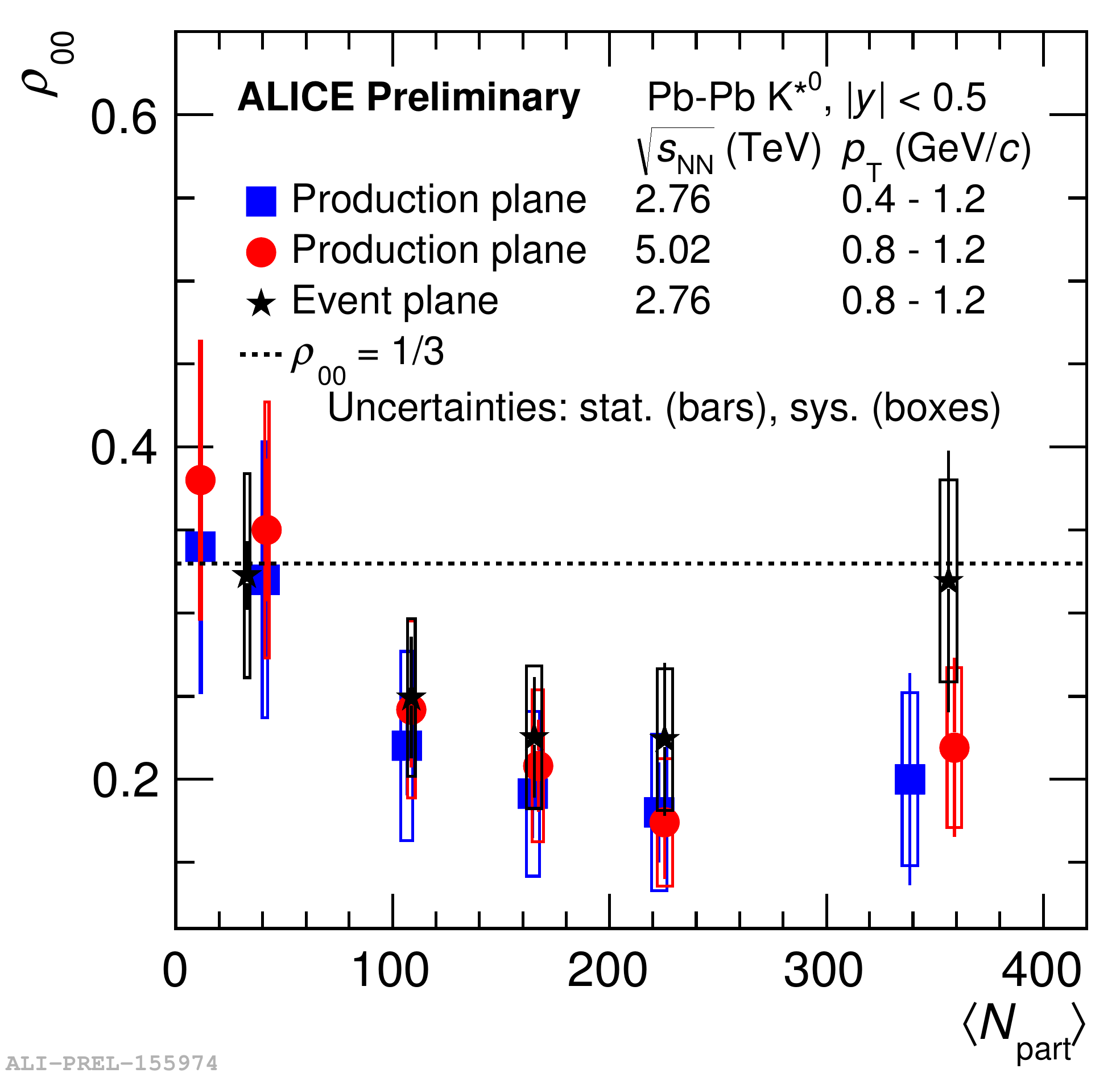}
\includegraphics[clip,scale=0.27]
{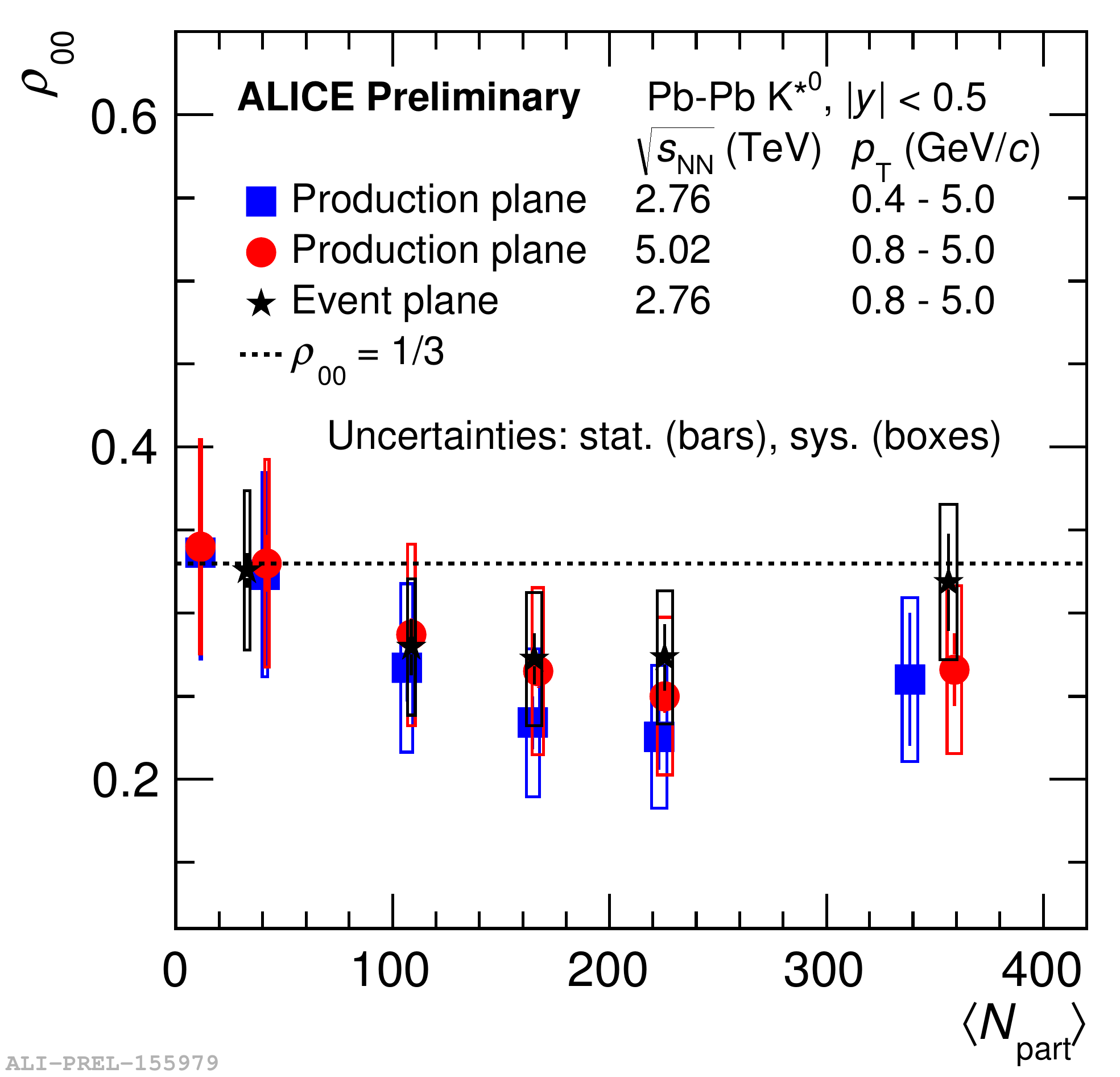}
\caption{(color online) Spin density matrix element
  \rh of K$^{*0}$ as a function of  $\langle N_{\mathrm{part}}\rangle$ w.r.t. production
  plane in Pb--Pb collisions at
  \snn = 2.76 TeV (blue marker) and 5.02 TeV (red marker)
  and w.r.t. event plane (black marker) in Pb--Pb collisions at \snn =
  2.76 TeV for the lowest \pt bin (left
  panel) and for the whole \pt range (right panel). The statistical uncertainties
  are shown as bars and systematic uncertainties are shown as
  boxes. The dotted line at \rh = 1/3
shows the no spin alignment scenario.}
\label{fig:centDep}
\end{center}
\end{figure}
\section{Results}
\label{sec:results} 
The left panel of Fig.~\ref{fig:ptDep} shows \rh as a function of
\pt for \kst in pp collisions at \s = 13 TeV, in 10-50\% Pb--Pb collisions at
\snn = 2.76 and 5.02 TeV  and \ks in
20-40\% Pb--Pb collisions at \snn = 2.76 TeV using the production
plane.  A comparison of \kst results using the production and event planes
in Pb--Pb collisions at \snn = 2.76 TeV is shown in the right panel of
Fig.~\ref{fig:ptDep}. The \kst results in pp collisions
and \ks (spin 0) in Pb--Pb collisions show no spin alignment (\rh
$\sim$ 1/3). The \rh values of \kst are lower than 1/3 at low \pt ($<$ 2.0 GeV/c)
for both production and event planes. To quantify the results, the \rh
value of \kst  for 0.4 $\leq$ \pt ~$<$ 1.2 GeV/$c$ is about
2.5$\sigma$ lower than 1/3 and about 2.3$\sigma$ lower
than 1/3 for  0.8 $\leq$ \pt ~$<$ 1.2 GeV/$c$ using the production plane
in Pb--Pb collisions at \snn = 2.76 and 5.02 TeV respectively.
The \rh value for \kst at  0.8 $\leq$ \pt ~$<$ 1.2GeV/$c$ using event
plane in Pb--Pb collisions at \snn = 2.76 TeV is about 1.7$\sigma$ lower than 1/3.
The \rh values for \kst are consistent
with 1/3 for higher \pt at both energies w.r.t. the production and
event planes.
Figure~\ref{fig:centDep} shows \rh as a function of $\langle
N_{\mathrm{part}}\rangle$ in Pb--Pb collisions for the lowest \pt bin 
used in this analysis (left panel) and for the whole \pt range (right panel). \rh values show a
clear centrality dependence in Pb--Pb collisions for both production
and event planes. The maximum deviation of \rh values from 1/3 is for
mid central (10-30\%) collisions that is expected due to the large
angular momentum for  semi central collisions. Within statistical
and systematic uncertainties the \rh values do not show energy
dependence and similar values are observed both for production and
event planes.

\section{Summary and outlook}
\label{sec:sum}
We have presented results on the spin alignment of \kst vector mesons
at midrapidity in Pb--Pb collisions at \snn = 2.76 and 5.02 TeV. The
spin alignment is measured as a function of \pt and collision
centrality classes w.r.t. the production and event planes. At low \pt
($<$ 2.0 GeV/$c$) the \rh values deviate from 1/3 for both the production
 and event planes. Within statistical and systematic uncertainties
the spin alignment results show no energy dependence. \rh shows a
clear centrality dependence for both the kninematic planes and the
maximum deviation of \rh from no spin alignment value 1/3 is for
mid-central (10-30\%) collisions.

The spin alignmnet studies of \kst w.r.t. the event plane in Pb--Pb
collisions at \snn = 5.02 TeV is ongoing. Furthermore, an increase in statistical precision is
expected with more data in Pb--Pb
collisions at \snn = 5.02 TeV.



\bibliographystyle{elsarticle-num}
\bibliography{<your-bib-database>}







\end{document}